\documentclass[11pt,fleqn,twoside]{article}

\usepackage{hamza2}
\makeatletter
\newcommand{\prava}{\footnotesize\it
\begin{flushright}
\begin{minipage}{18cm}
Copyright \copyright 1998 by A.M. Hamza
\end{minipage}
\end{flushright}}

\newcommand{\name}[1]{\begin{flushleft}
                       \LARGE \bf #1
                       \end{flushleft}\vspace{-3mm}}

\newcommand{\Author}[1]{\begin{flushleft}
                       \it #1 \end{flushleft}}

\newcommand{\Adress}[1]{\begin{flushleft}
                       \it #1 \end{flushleft}}

\newcommand{\Date}[1]{\begin{flushleft}
                      \small  \it #1 \end{flushleft}}

\newcommand{\ehkol}{Author \ name}
\newcommand{\ohkol}{Article \ name}
\renewcommand{\@evenhead}{
\hspace*{-3pt}\raisebox{-15pt}[\headheight][0pt]{\vbox{\hbox to \textwidth
{\thepage \hfil \ehkol}\vskip4pt \hrule}}}
\renewcommand{\@oddhead}{
\hspace*{-3pt}\raisebox{-15pt}[\headheight][0pt]{\vbox{\hbox to \textwidth
{\ohkol \hfil \thepage}\vskip4pt\hrule}}}
\renewcommand{\@evenfoot}{}
\renewcommand{\@oddfoot}{}

\newcommand{\be}{\begin{equation}}
\newcommand{\ee}{\end{equation}}
\newcommand{\ba}{\hspace*{-5pt}\begin{array}}
\newcommand{\ea}{\end{array}}

\newcommand{\ds}{\displaystyle}
\makeatother

\begin{document}
\setcounter{page}{462}

\thispagestyle{empty}

\renewcommand{\ehkol}{A.M. Hamza}

\renewcommand{\ohkol}{Mode-Coupling and Nonlinear Landau Damping Ef\/fects}

\begin{flushleft}

\footnotesize \sf

Journal of Nonlinear Mathematical Physics \qquad 1998, V.5, N~4,
\pageref{hamza2-fp}--\pageref{hamza2-lp}.
\hfill {\sc Article}

\end{flushleft}

\vspace{-5mm}

\renewcommand{\footnoterule}{}

{\renewcommand{\thefootnote}{} \footnote{\prava}

\name{Mode-Coupling and Nonlinear Landau Damping Ef\/fects in Auroral
Farley-Buneman Turbulence}\label{hamza2-fp}

\Author{A.M. HAMZA}
\Adress{Department of Physics, Center for Space Research, University of New
Brunswick, Fredericton, NB, E3A-5A3, Canada}

\Date{Received June 2, 1998; Accepted September 01, 1998}

\begin{abstract}

\noindent
The fundamental problem of Farley-Buneman turbulence in the auroral
$E$-region has been discussed and debated extensively in the past two 
decades. In
the present paper we intend to clarify the dif\/ferent steps that the auroral
$E$-region plasma has to undergo before reaching a steady state. The
mode-coupling calculation, for Farley-Buneman turbulence, is
developed in order to place it in perspective and to estimate its
magnitude relative to the anomalous ef\/fects which arise through the
nonlinear wave-particle interaction. This nonlinear ef\/fect, known as
nonlinear ``Landau damping'' is due to the coupling of waves which
produces other waves which in turn lose energy to the bulk of the
particles by Landau damping. This leads to a decay of the wave energy
and consequently a heating of the plasma. An equation governing the
evolution of the f\/ield spectrum is derived and a physical
interpration for each of its terms is provided.
\end{abstract}

\section{Introduction}

The selfconsistent theory of Farley-Buneman turbulence as developed by
{\it Sudan}~[8, 9], and by {\it Hamza and St-Maurice} [3, 4] has
addressed successfully a number of features observed by radars in the
equatorial and auroral ionosphere. Recently, {\it Hamza and
St-Maurice}~[5, 6] developed a general f\/luid formalism, taking into
account the previously ignored nonlinear electron inertia ef\/fect term. The
new turbulence theory provides a possible explanation for the existence of
large aspect angle echoes in the auroral $E$-region, and justif\/ies
selfconsistently the link between electron heating along the magnetic f\/ield
lines and the anomalous ef\/fects induced by perpendicular Farley-Buneman
turbulence.  In [5, 6] it is shown that the advection of parallel current by
perpendicular drifts leads to an anomalous parallel collision frequency. In
other words, it is shown that perpendicular Farley-Buneman turbulence can
indeed lead to anomalous parallel ef\/fects which in turn allow for parallel
electric f\/ields to be generated. This has an impact on energy conservation
since the zeroth order Ohmic heating which was conf\/ined to the plane
perpendicular to the magnetic f\/ield, can now have a parallel component due
to plasma wave turbulence $\langle J_{e\parallel} E_{\parallel}\rangle$. This
is made possible by the nonlinear electron inertia term. In order to clarify
the physics related to the energy transport properties of the $E$-region
plasma we have to discuss the dif\/ferent mechanisms of energy
tranfer from the free energy source conf\/ined to the plane
perpendicular to the magnetic f\/ield to the particles moving along
the f\/ield lines. This can only be achieved by 
including mode-coupling ef\/fects and describing their role as it
pertains to the f\/inal picture.

In what follows, we will use the results of {\it Hamza and
St-Maurice} [5, 6] to establish the f\/inal link by identifying the
mode-coupling terms and the terms associated with the analog of a
wave-particle interaction respectively. In particular, we will stress
the fact that the mode-coupling terms are energy conserving and that
they do contribute a frequency broadening $\Delta\omega_{\bf k}$ as
already shown by {\it Hamza and St-Maurice}~[3] as well as 
{\it Sudan}~[8, 9]. Finally we will discuss the
roles of the mode-coupling terms and that of the anomalous ef\/fects
generated via nonlinear wave-particle analogs respectively. However, more
important than the comparison of the relative magnitudes of the mode-coupling
rate and the parallel rate of dissipation generated by the anomalous
ef\/fects is the dif\/ference in nature of the physics associated
with each respectively. Although, the frequency broadening generated
by mode-coupling, as shown by {\it Hamza and St-Maurice} [3], might
indeed balance the linear growth rate of a Farley-Buneman wave, it is
only shuf\/f\/ling energy into other waves. The parallel dissipation
rate, on the other hand, allows wave energy to be channelled 
directly into electron kinetic energy. In steady state turbulence a physical
process must ultimately convert wave energy into thermal energy. This energy
conservation, in fact is a requirement for any selfconsistent theoretical
model.

We already know that during the so-called ``quasi-linear'' phase, the
background particle distribution dif\/fuses in such a way as to bring
the growth rate of the unstable waves to zero (``plateauing'' of the
distribution function), leaving a quasi-stationary spectrum interact
in such a way as to distort the spectrum but keep the energy in the
spectrum roughly constant. This is achieved by extracting energy from
the resonant particles and dumping it unto the waves. 
Quasi-linear theory leads to the development of a quasi-equilibrium spectrum
which persists indef\/initely. The spectral energy can be shown to be
proportional to $\gamma/\omega$, where $\gamma$ and $\omega$ represent the
growth rate and the eigenfrequency respectively. This energy is small since
it is proportional to $\gamma/\omega\ll 1$, and therefore can not be
considered to be an ultimate saturation mechanism for the problem at
hand as shown by {\it St-Maurice} [7]. Mode-coupling ef\/fects,
neglected by the quasi-linear theory lead to a broadening of the
equlibrium spectrum. The mode-coupling ef\/fects can be
divided into two categories. The f\/irst class of terms can be qualif\/ied as
``resonant'' in the sense that waves couple while satisfying the following
relations (frequency and wavemomentum matching)
\be
\ba{l}
\omega_{{\bf k}^{'}}+\omega_{{\bf k}^{''}} =  \omega_{\bf k},
\vspace{1mm} \\
{\bf k}^{'}+{\bf k}^{''}  =  {\bf k}. 
\ea\label{hamza2_eq:i1}
\ee
The resonant mode coupling terms lead to a broadening of the
quasi-linear spectrum and the time scale associated with this
ef\/fect is proportional to the spectral density (in our case
$|\delta\phi|^{2}\propto\gamma$), and therefore the
resonant mode coupling process is on the same time scale as the quasi-linear
time scale. However, as shown by {\it Hamza and St-Maurice}~[3, 4] and {\it
Drummond and Pines} [2] before them, the resonant mode coupling leaves the
total energy in the wave spectrum virtually untouched (constant).

One major constraint in resonant mode coupling is expressed by equation
(\ref{hamza2_eq:i1}). The driving term is a quadratic nonlinearity of the form
$\phi_{{\bf k^{'}}}(t)\Phi_{{\bf k}-{\bf k^{'}}}(t)$ and induces a
time dependence of $\exp{[i(\omega_{{\bf k^{'}}}+\omega_{{\bf k}-{\bf
k^{'}}})t]}$, which in general can be very dif\/ferent from the time
dependence of $\phi_{{\bf k}}(t)$, that is, $\exp{[i\omega_{\bf k}
t]}$. In other words, in the general case one has 
\be
\ba{l}
\ds \omega_{{\bf k}^{'}}+\omega_{{\bf k}^{''}} \neq \omega_{\bf k},
\vspace{1mm} \\ 
 {\bf k}^{'}+{\bf k}^{''}    =   {\bf k}.
\ea
 \label{hamza2_eq:i2}
\ee
The nonlinear source forces the f\/ield $\phi_{\bf k}(t)$ to have
terms with the time dependence of the source rather than the natural
time dependence of the f\/ield itself. This is very much the case for
a forced harmonic oscillator. This leads to separating the f\/ield
variable into two parts. The f\/irst part gives rise
to resonant coupling because of frequency matching, while the second part
arises because of the non-resonant parts of the nonlinear source. An essential
feature of the latter part is that $(\omega_{\bf k^{'}}+\omega_{{\bf
k}^{''}})$ can be fairly small so as to allow the phase velocity of
the resultant wave $(\omega_{\bf k^{'}}+
\omega_{{\bf k}-{\bf k}^{'}})/k_{\parallel}$ to be smaller than the
electron thermal velocity, and for
that matter the ion thermal velocity. This means that the non-resonant
contribution of the mode-coupling terms can undergo Landau damping. The
non-resonant waves Landau damp and in the process give their energy
to both the electron and ions, i.e. heat the plasma along the f\/ield lines.

The main objective of the present calculation, is to
show that the nonlinearities in Farley-Buneman turbulence contribute
coherent and resonant terms. These terms are isolated and shown to
have dif\/ferent impacts on the physics, especially on the energy
conservation constraint. We have omitted to discuss transient
phenomena because we are only interested in the global energy
conservation on a long time scale. Higher order nonlinearities have
also been neglected in the perturbation analysis. The model has been
extensively described in~[6]. Therefore we will refer the reader to
that paper for formal manipulations and for details related to the
derivation of certain fundamental results.

\section{Theoretical Model}

The fundamental assumptions and the details of the theoretical model are
described, as mentioned above in~[6]. We will only quote the results
neccessary for the development of an energy balance equation
containing all the relevant terms including the mode-coupling terms.

The starting point of this paper is equation (24) of~[6] which can be
written in a compact form as follows
\begin{equation}
-in_0\frac{e\phi_{{\bf k}\omega}}{T_e}\left(\omega-\omega_{\bf
k}^{(L)}\right)=(1)+(2)+(3),
\label{hamza2_eq:1a} 
\end{equation}
where the three terms on the right hand side of equation
(\ref{hamza2_eq:1a}) can be explicitly written as
\begin{equation}
(1)\equiv-\frac{i}{1+\psikom}\frac{\mi}{e^{2}}
\frac{\om^{2}-\kperpsq\frac{\Ti}{\mi}+i\nui\om}{\kperpsq}\frac{\kpara}
{\Lambdak}\sumkomp\Akkps\frac{e\phikomp}{\Te}\Jeparakk. \label{hamza2_eq:1b}
\end{equation}
This term represents the advection of parallel current by perpendicular and
parallel drifts through electron inertia. The second term corresponds to
the  advection of perpendicular current by parallel and perpendicular
drifts and can be written in the following form
\be
\ba{l}
\ds
(2)\equiv\frac{i}{1+\psikom}\frac{\mi}{e^{2}}\frac{\om^{2}-\kperpsq\frac{\Ti}
{\mi}+i\nui\om}{\kperpsq}\Omeomtilde
\vspace{3mm}\\
\ds \qquad \times \sumkomp\left(\frac{\Ome}{\Lambdakkp}-\betae \right)\Akkps
\frac{e\phikomp}{\Te}{\bf k^{'}}erp\cdot\bfJeperpkk.
\ea
\label{hamza2_eq:1c}
\ee
It is important to emphasize the fact that these two terms were completely
omitted by previous theories addressing the issues related to Farley-Buneman
turbulence in the auroral as well as equatorial $E$ region.

Finally the last term represents the classical mode coupling terms,
which can be written in the following form
\be
\ba{l}
\ds
(3)\equiv-\frac{\no}{1+\psikom}\Omeomtilde\sumkomp\bkkps\akkptilde
\vspace{3mm}\\
\ds \qquad \times \vtesq\frac{e\phikomp}{\Te}\frac{e\phikk}{\Te}
-\no\frac{\Ome}{1+\psikom}\Omeomtildesq
\vspace{3mm}\\
\ds \qquad \times \sumkomp\frac{\Akkps}{\Lambdakkp}\bkkps
\ckkps\frac{\zhat\cdot({\bf k}\times{\bf k^{'}})}{|{\bf
k}_\perp-{\bf k^{'}}_\perp|}\frac{e\phikomp} 
{\Te}\phikk.
\ea \hspace{-13.63pt} \label{hamza2_eq:1}
\ee
In the above expressions we have
\begin{equation}
\psikom=\psikoms \label{hamza2_eq:2}
\end{equation}
while $\omkL$ represents the linear eigenfrequency for Farley-Buneman
waves. The coef\-f\/i\-cients $\Akkps$, $\bkkps$, and $\ckkps$ are
def\/ined in the paper~[6], and the parallel and perpendicular
currents $J_{e}^{\parallel}$ and ${\bf J}_{e}^{\perp}$ are given by
equations (21) and (22) in~[6] respectively. 

It is clear from equation (\ref{hamza2_eq:1})
that when the nonlinear right hand side
is neglected we recover the classic Farley-Buneman dispersion relation.

The f\/irst two terms on the right hand side of equation
(\ref{hamza2_eq:1a}) contribute the coherent terms discussed extensively in
the paper~[6]. The coherent terms 
can easily be identif\/ied through a formal substitution of the
expressions for $J_{e}^{\parallel}$ and ${\bf J}_{e}^{\perp}$ into
equation (\ref{hamza2_eq:1a}). 
Isolating these coherent terms, while neglecting the four-wave coupling which
generates cubic nonlinearities, allows us to rewrite the top equation
(\ref{hamza2_eq:1}) in the following form. To illustrate this iteration process let
us consider the f\/irst term in equation (\ref{hamza2_eq:1a}). To proceed we
need the expression for the parallel current
\begin{equation}
\ba{l}
\ds \Jeparakom=-i\fracnoesqme\frac{\kpara}{\Lambdak}\Fkoms\phikom
-\sumkomp\frac{\Akkps}{\Lambdak}\phikomp\Jeparakk
\vspace{3mm}\\
\ds \qquad 
-i\frac{\esq}{\me}\sumkomp\frac{\kparap}{\Lambdak}
\phikomp\deltankk.
 \label{hamza2_eq:2a}
\ea \end{equation}
Substituting equation (\ref{hamza2_eq:2a}) into
equation (\ref{hamza2_eq:1b}) one
obtains two types of terms. The f\/irst type can be combined with the
mode coupling term given
by equation (\ref{hamza2_eq:1c}), while the second type of term, a cubic term, is of
the form $\phi\phi J_{e}^{\parallel}$. A second iteration of this
term leads to a cubic term of the form $|\phikomp|^{2}\phikom$. This
term can combined with the left hand side of equation (\ref{hamza2_eq:1a})
to obtain a renormalized dispersion relation. This ef\/fect
introduces an anomalous frequency solely due to nonlinear 
ef\/fects analog to a higher order wave-particle interaction.
\begin{equation}
-i(\om-\omk)\phikom=\sumkomp\Mkkps\phikomp\phikk, \label{hamza2_eq:3}
\end{equation}
where $\Mkkps$ is the nonlinear coupling coef\/f\/icient obtained by
substituting the expressions for $J_{e}^{\parallel}$ and ${\bf
J}_{e}^{\perp}$, as given in~[6], in equation (\ref{hamza2_eq:1a}) above.
Where the new frequency $\omk$ includes 
the coherent terms generated  by the nonlinearities, and written as
$\nu^{\parallel}J_{e}^{\parallel}$ and $\nu^{\perp}{\bf
J}_{e}^{\perp}$, where $J_{e}^{\parallel}$ and ${\bf J}_{e}^{\perp}$
are shown to be propotional to $\phikom$ to f\/irst order. These
nonlinear coherent expressions were derived and 
discussed extensively in~[6]. $\omk$ can now be written as follows:
\begin{equation}
\omk=\omkL+ \omk^{C},\label{hamza2_eq:4a}
\end{equation}
where $\omk^{C}$ represents the anomalous (coherent) contribution by
the nonlinear wave-particle analogs in equation (\ref{hamza2_eq:1}). As one
can expect, the nonlinear expression for this anomalous frequency is
quite complicated and will not be given here.

Making the following change in wavevectors and frequencies ${\bf
k^{'}}\rightarrow {\bf k}-{\bf k^{'}}$, and $\omp\rightarrow\om-\omp$
leads to rewriting equation (\ref{hamza2_eq:3}) in a useful form for our
purpose as will become clear in what follows.
\begin{equation}
-i(\om-\omk)\phikom=\frac{1}{2}\sumkomp\left(\Mkkps+\Mkmkps\right)
\phikomp\phikk.
\label{hamza2_eq:4}
\end{equation}
Multiplying by $\phikomstar$ and ensemble averaging leads to the evolution
equation for the spectral density $\spectrekom$
\begin{equation}
-i(\om-\omk)\spectrekom=\frac{1}{2}\sumkomp\left(\Mkkps+\Mkmkps\right)
\left\langle\phikomp\phikk\phikomstar\right\rangle.
\label{hamza2_eq:5}
\end{equation}
At this stage one has to close the system by prescribing the triple
correlation function. We shall assume that the distribution of potential
f\/luctuations is Gaussian, which allows to evaluate explicitly, via a
perturbation scheme, the right hand side of equation (\ref{hamza2_eq:5}), that is
\[
\phi=\phi^{(1)}+\phi^{(2)}+\cdots
\]
with
\be
\ba{l}
\ds -i\left(\om-\omkL\right)\phikom^{(1)}=0,
\vspace{3mm}\\
\ds -i(\om-\omk)\phikom^{(2)}= \frac{1}{2}\sumkomp\left(\Mkkps+\Mkmkps\right)
\phikomp^{(1)}\phikk^{(1)} 
\ea \label{hamza2_eq:6}
\ee
the triple correlation function can be written as:
\be
\ba{l}
\ds \left\langle\phikomp\phikk\phikomstar\right\rangle
=  \left\langle\phikompo
\phikko\phikomostar\right\rangle  
+ \left\langle\phikompo \phikko\phikomtstar\right\rangle
\vspace{3mm}\\ 
\ds \qquad  +  \left\langle\phikompo
\phikkt\phikomostar\right\rangle   +  \left\langle\phikompt
\phikko\phikomostar\right\rangle.
\ea \label{hamza2_eq:7}
\ee
The f\/irst term on the right hand side of equation (\ref{hamza2_eq:7})
vanishes, while the others can be combined with equation (\ref{hamza2_eq:6})
to lead to 
\be
\ba{l}
\ds \left\langle\phikompo\phikko\phikomtstar\right\rangle  = 
-\frac{i}{2}\sumkomp
\frac{\left(\Mkkps+\Mkmkps\right)^{*}}{(\om-\omk)}\spectrekomp\spectrekk,
\vspace{3mm} \\
\ds \left\langle\phikompo\phikkt\phikomostar\right\rangle  = 
\frac{i}{2}\sumkomp\frac{\left(\Mkkmkps+\Mkkks\right)}
{(\om-\omp-\omkkp)}\spectrekomp\spectrekom,
\vspace{3mm}\\
\ds \left\langle\phikompt\phikko\phikomostar\right\rangle  = 
\frac{i}{2}\sumkomp\frac{\left(\Mkpkkps+\Mkpks\right)}
{(\omp-\omkp)}\spectrekomp\spectrekom.
\ea \hspace{-5.53pt}\label{hamza2_eq:8}
\ee

This f\/inally allows us to write the equation (\ref{hamza2_eq:5}) for the
spectral density in the following form:
\be
\ba{l}
\ds -i(\om-\omk)\spectrekom  = 
\frac{1}{2}\sumkomp\left(\Mkkps+\Mkmkps\right)
\vspace{3mm}\\
\ds \qquad \times \left\{
\frac{i}{(\om-\omp-\omkkp)}\left(\Mkkmkps
+ \Mkkks\right)\spectrekomp\spectrekom\right.
\vspace{3mm} \\
\ds \qquad  - \left.\frac{1}{2}\frac{i}{(\om-\omk)}\left
(\Mkkps+\Mkmkps\right)^{*}\spectrekomp\spectrekk\right\}.
\ea \label{hamza2_eq:9}
\ee
The f\/irst term on the right hand side corresponds to a broadening
due to resonant three-wave coupling (mode coupling), and can be
incorporated with the left hand side to give
\be
\ba{l}
\ds -i(\om-\omk-\Delta\omk)\spectrekom
\vspace{3mm}\\
\ds \qquad =-\frac{1}{4}\frac{i}{(\om-\omk)}\sumkomp
\left(\Mkkps+\Mkmkps\right)^{*}
\spectrekomp\spectrekk,
\ea \label{hamza2_eq:10}
\ee
where
\begin{equation}
\Delta\omk=-\frac{1}{2}\sumkomp\frac{\left(\Mkkps+\Mkmkps\right)
\left(\Mkkmkps+\Mkkks\right)}{(\om-\omp-\omkkp)}\spectrekomp.
\label{hamza2_eq:11}
\end{equation}
These results are very much similar to the two dimensional results
published by {\it Hamza and St-Maurice}~[4] when looking
explicitly at the two dimensional Farley-Buneman turbulence. The
coef\/f\/icients in the current case are more complicated because we
have taken into account the electron inertia nonlinearities and
included the parallel ef\/fects. We do not intend to explicitly write
the expressions for the coef\/f\/icients since, as might be 
expected they are quite complicated and would not serve the purpose of this
paper. However, we have in equation (\ref{hamza2_eq:10}) isolated two types of terms.
The f\/irst term is a coherent analog of a nonlinear wave-particle
interaction, nonlinear Landau-damping, while the second is due to a resonant
three-wave coupling. The term on the right hand side of equation
(\ref{hamza2_eq:10}) is due solely to mode coupling.

At this stage, one can study two limiting cases, namely a strong turbulence
limit and a weak tubulence one.

\subsection{Strong Turbulence Limit}
In this case equation (\ref{hamza2_eq:10}) can be written in a more familiar form:
\begin{equation}
\!\!\spectrekom=\frac{1}{4}\frac{1}{(\om-\omk-\Delta\omk)^{2}}\sumkomp
\left(\Mkkps+\Mkmkps\right)^{*}\!\!\spectrekomp\spectrekk. \label{hamza2_eq:12}
\end{equation}
This equation can be written in the following form
\begin{equation}
\spectrekom=\frac{\langle
|{\tilde\phi}_{{\bf k}\om}|^{2}\rangle}{|\epsilon_{{\bf k}\om}|^{2}},
\label{hamza2_eq:12a}
\end{equation}
where $\langle |\tilde\phi_{{\bf k}\om}|^2\rangle$ is the
incoherent driven potential obtained by identifying the right hand
side of equation (\ref{hamza2_eq:12a}) to the right hand side of equation
(\ref{hamza2_eq:12}). This has been the object of a long debated issue
related to describing strong turbulence as it occurs in plasmas. We
will try to illustrate, very brief\/ly, the results of the debate on
strong turbulence using the fundamental results of {\it Boutros-Ghali and Dupree}~[1] and references therein. When writing 
Poisson's equation we separate the coherent (phase coherent with the
electric f\/ield) from the incoherent contribution of the
distribution function in the following way 
\begin{equation}
{\bf\nabla}\cdot\delta{\bf E}=4\pi nq\int d{\bf v}\left(\delta
f^{c}+\delta{\tilde f}\right), \label{hamza2_eq:12b}
\end{equation}
where $\delta f^{c}$ and $\delta{\tilde f}$ represent the coherent  and
incoherent parts of the particle distribution function respectively. The
coherent part contributes to the dispersion relation which allows us
to write 
\begin{equation}
i\epsilon_{{\bf k}\om}{\bf k}\cdot\delta{\bf E}=4\pi nq\int d{\bf
v}\delta{\tilde f} 
\label{hamza2_eq:12c}
\end{equation}
using $\delta {\bf E}=-i{\bf k}\phikom$ leads to
\begin{equation}
\phikom=\frac{4\pi nq}{k^{2}\epsilon_{{\bf k}\om}}\int d{\bf v}\delta{\tilde
f}=\frac{\tilde \phi_{{\bf k}\om}}{\epsilon_{{\bf k}\om}} \label{hamza2_eq:12d}
\end{equation}
which basically leads  to equation (\ref{hamza2_eq:12a}). A strong turbulence result
(see for example {\em Boutros-Ghali and Dupree}~[1]).

\subsection{Weak Turbulence Case}

In the weak turbulence regime, in which the waves are weakly growing,
with small line widths $Im \,(\omk)$, and replacing Fourier series with
Fourier integrals, we obtain
\begin{equation}
\sumkom\spectrekom\rightarrow\int\frac{d\bf k}{(2\pi)^{4}}\int
d\om\langle\phi^{2}\rangle_{{\bf k}\om} \label{hamza2_eq:13}
\end{equation}
and using the following spectral relation
\begin{equation}
\langle\phi^{2}\rangle_{{\bf k}\om}=\langle\phi^{2}\rangle_{{\bf k}}2\pi
\delta(\om-\omk). \label{hamza2_eq:14}
\end{equation}
Let us now go back to equation (\ref{hamza2_eq:10}) and
integrate over the frequency
using the following identity
\begin{equation}
-2Re \int\frac{d\om}{2\pi}i(\om-\omkL)\spectrekom=\frac{d}{dt}
\langle\phi^{2}\rangle_{{\bf k}}. \label{hamza2_eq:15}
\end{equation}
In other words, the time derivative represents the quasi-linear
growth rate. We f\/inally obtain
\be
\ba{l}
\ds \frac{d}{dt}\langle\phi^{2}\rangle_{{\bf k}}= -2Re
\int\frac{d\om}{2\pi}i(\omk^{C}+\Delta\omk)\spectrekom
\vspace{3mm}\\
\ds \qquad - 2Re \int\frac{d{\bf k^{'}}}{(2\pi)^{3}}\left(M_{{\bf
k};{\bf k^{'}}} +M_{{\bf k};{\bf k}-{\bf k^{'}}}\right)^{*}
\langle\phi^{2}\rangle_{{\bf k^{'}}}\langle\phi^{2}\rangle_{{\bf k}-
{\bf k^{'}}}\pi \delta(\omk-\omkp-\om_{{\bf k}-{\bf k^{'}}})
\ea \label{hamza2_eq:16}
\ee
the f\/irst two terms contribute two dif\/ferent growth rates:
$\gamma_{{\bf k}}^{C}$ due to the coherent nonlinear wave-particle analog, and
$\gamma_{{\bf k}}^{\Delta}$ due to mode coupling ef\/fects. This
allows us to write the following spectral equation
\be
\ba{l}
\ds \frac{d}{dt}\langle\phi^{2}\rangle_{{\bf k}}=-2(\gamma_{{\bf k}}^{C}+
\gamma_{{\bf k}}^{\Delta})\langle\phi^{2}\rangle_{{\bf k}}
\vspace{3mm}\\
\ds \qquad - 2Re\int\frac{d{\bf k^{'}}}{(2\pi)^{3}}\left(M_{{\bf
k};{\bf k^{'}}} +M_{{\bf k};{\bf k}-{\bf k^{'}}}\right)^{*}
\langle\phi^{2}\rangle_{{\bf k^{'}}}
\langle\phi^{2}\rangle_{{\bf k}-{\bf k^{'}}}\pi
\delta(\omk-\omkp-\om_{{\bf k}-{\bf k^{'}}})
\ea \label{hamza2_eq:17}
\end{equation}
integrating over ${\bf k}$ one can show that the
mode-coupling term conserves the total mode energy, which leaves us with:
\begin{equation}
\int d{\bf k}\frac{d}{dt}\langle\phi^{2}\rangle_{{\bf k}}=-2\int
d{\bf k} (\gamma_{{\bf k}}^{C})\langle\phi^{2}\rangle_{{\bf k}}
\label{hamza2_eq:18}
\end{equation}
in other words, in order to reach a steady state the right hand side has to
vanish. This leads us to the following interpretation. A steady state requires
both resonant mode coupling and the analog of a wave-particle exchange to take
place; one in the absence of the other does not neccessarily guarantee steady
state turbulence. Consequently, a steady state in Farley-Buneman turbulence
requires that the free energy available in the plane perpendicular to the
magnetic f\/ield, extracted via a linear instability, be converted
into particle kinetic energy along the f\/ield lines via an analog of
a nonlinear wave-particle mechanism.

\section{Summary and Conclusions}

In this letter we have established the neccessary condition for steady state
Farley-Buneman turbulence. We have shown that in order to reach steady state
the growth in the plane perpendicular to the magnetic f\/ield has to be
compensated by a growth along the f\/ield line; A perfect analogy with a leaky
bucket, if the we want the water level in the bucket to remain unchanged then
we need to balance the rate of pouring to the rate of leaking water. In other
words free energy is extracted from the drifts by the Farley-Buneman waves
which start to grow and couple. This process will go on, and no steady state
will be reached unless the energy is given back to the particles, in this case
the particles moving along the f\/ield lines. So two processes take place. First,
linear growth and resonant coupling of Farley-Buneman waves, then a nonlinear
wave particle coupling along the f\/ield lines. These two mechanisms when
combined together can lead to steady state turbulence.

\subsection*{Acknowledgments}

The author is grateful to J.-P. St-Maurice and D.R.~Moorcroft for
very useful discussions, and thanks D.~Montgomery for the private
communication. Funding for this research has been provided by the
Canadian Space Agency.

\label{hamza2-lp}

\end{document}